\documentclass[reprint, amsmath,amssymb, aps,prl]{revtex4-1}

\usepackage{graphicx}% Include figure files
\usepackage{dcolumn}% Align table columns on decimal point
\usepackage{bm}% bold math
\usepackage{color} % for \textcolor
\usepackage{gensymb} % for \degree
\usepackage{amsmath} % for equation typesetting
\usepackage[T1]{fontenc} % for text arrows 1
\usepackage[utf8]{inputenc} % for text arrows 2
\usepackage{textcomp} % for text arrows 3

\newcommand{\negsp}{ % for "tightening" long equations
        \!\!\!\!\!\!
}

\begin{document}

\title{A protrusion can ``eclipse'' looping of a long self-avoiding chain}
\author{Yaroslav Pollak}
\author{Sarah Goldberg}
\author{Roee Amit}
\email{roeeamit@technion.ac.il}
\affiliation{%
        Biotechnology and Food Engineering and Russell Berrie Nanotechnology Institute, Technion – Israel Institute of Technology, Haifa, Israel 32000
}

\date{\today}

\begin{abstract}
        We simulate long self-avoiding chains using a weighted-biased sampling  Monte-Carlo algorithm, and compute the probabilities for chain looping with and without a protrusion. We find that a protrusion near one of the chain's termini reduces the probability of looping, even for chains much longer than the protrusion--chain-terminus distance. This effect increases with protrusion size, and decreases with protrusion-terminus distance. We model the simulated results theoretically by considering how the protrusion ``eclipses'' the chain terminus closer to the protrusion from the more distant chain terminus. This eclipse mechanism has implications for understanding the regulatory role of proteins bound to DNA.
\end{abstract}

\maketitle

Polymer looping is a phenomenon that is critical for the understanding of many chemical and biological processes. In particular, DNA looping has been implicated in transcriptional regulation across many organisms, and as a result plays a crucial role in how organisms develop and respond to their environments. While DNA looping has been studied extensively over the last several decades both experimentally \cite{marko_stretching_1995,Mueuller1996,cloutier_dna_2005,wiggins_high_2006,vafabakhsh_extreme_2012}
and theoretically \cite{jacobson_intramolecular_1950,flory_statistical_1969,gennes_scaling_1979,shimada_ring-closure_1984,crothers_dna_1992,yan_statistics_2005},
many aspects of looping-based transcriptional regulation remain poorly understood.

In this Letter we address the question of how a protrusion affects looping of a long polymer, with an emphasis on understanding experimentally-observed phenomena~\cite{Arnosti1996, Gray1996}. We use a modified worm-like chain model that takes excluded-volume considerations into account \cite{Pollak2014,chen_renormalization-group_1992,tree_is_2013}.
We show that the excluded volumes of a polymer and of an object bound to it can block the ``line-of-sight'' of the two distant termini of the polymer, which in turn leads to a reduction in the probability of looping.

\emph{Polymer in the absence of bound objects.} The polymer is modeled as a discrete semi-flexible chain made of individual links of length $l$. A chain is described by the locations $\mathbf{r}_{i}$
of its link ends, and a local coordinate system defined by three orthonormal vectors $\hat{u}_{i}$, $\hat{v}_{i}$, $\hat{t}_{i}$
at each link, where $\hat{t}_{i}$
points along the direction of the $i$th link. We use the following notations for a specific chain configuration: $\theta_{i}$, $\phi_{i}$
are the zenith and azimuthal angles of $\hat{t}_{i}$
in local spherical coordinates of link $i-1$,
respectively. $\{\theta,\phi\}_{n}\equiv\{\theta_{1},...,\theta_{n},\phi_{1},...,\phi_{n}\}$
denotes all the angles until link $n$. Joint $i$
is the end-point of link $i$ and  joint $0$ is the beginning terminus of the chain. $w$
is the effective cross-section of the polymer. Each chain joint is engulfed by a “hard-wall” spherical shell of diameter $w$. The total elastic energy associated with the polymer chain can be written as follows \cite{Pollak2014}:
\begin{equation}
E\left(\left\{ \theta,\phi\right\}_{N} \right)=\sum_{i=2}^{N}E^{\mathrm{bend}}\left(\theta_{i},\phi_{i}\right)+\sum_{i=2}^{N}E_{i}^{\mathrm{hw}}\left(\left\{ \theta,\phi\right\}_{i} \right), \label{eq:SAWLC_E-1-1}
\end{equation}
where the elastic contribution to the energy is given by:
\begin{equation}
\beta E^{\mathrm{bend}}\left(\theta_{i},\phi_{i}\right)=a\left(1-\cos\theta_{i}\right), \label{eq:WLC_Bend_E}
\end{equation}
$a$ is the bending constant of the polymer chain, and we have assumed azimuthal symmetry. The hard-wall contribution is given by:
\begin{eqnarray}
\beta E_{i}^{\mathrm{hw}}\left(\left\{ \theta,\phi\right\}_{i} \right)=
\begin{cases}
\infty & i \text{ overlaps with one or more} \\
& \text{joints }0\ldots\left(i-\Delta i\right)\,\,.\\
0 & \text{otherwise}
\end{cases}\label{eq:E-hw}
\end{eqnarray}
Here $\beta=\left(k_{b}T\right)^{-1}$, $k_{b}$ is the Boltzmann factor and $T$ is the temperature. In case $l\geq w$, $\Delta i=1$. In case $l<w$, two or more consecutive spheres overlap and $\Delta i$
ensures that links $j$ and $k$ interact only if $\left|j-k\right|\geq\Delta i\geq\frac{w}{l}$. For simplicity, we disregard the twist degree of freedom in this work.

\begin{figure}[t]
        \centering{}
        \includegraphics[width=\columnwidth]{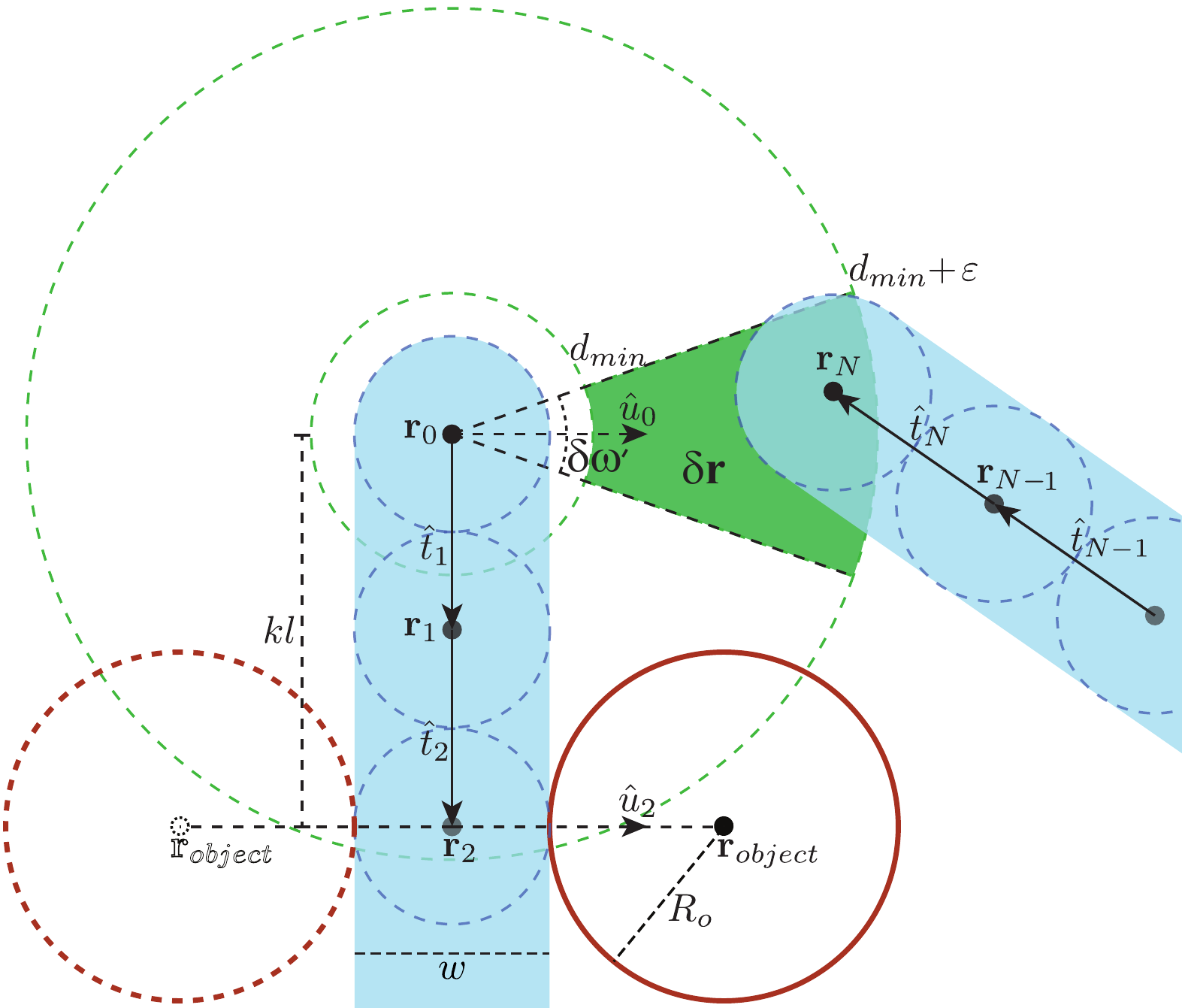}
        \caption{\label{fig:illustration} Loop and protrusion geometry. The chain (shaded in cyan) is modeled by spheres (blue dashed circles), here with link length $l$ equal to diameter $w$. Looping volume $\delta\mathbf{r}$ is represented by a green wedge. The protrusion is positioned on the second link ($k=2$). In-phase-with-$\delta\mathbf{r}$ ($\gamma_{2}=0\degree$) and out-of phase ($\gamma_{2}=180\degree$) positions are illustrated by the solid and dashed red circles, respectively.}
\end{figure}

\emph{Polymer in the presence of bound objects.} We model the bound objects as hard-wall spherical protrusions positioned adjacent to the polymer chain, with radius $R_{o}$ representative of the protrusion's volume (Fig.~\ref{fig:illustration}). Since we neglect torsion effects in our present model, there is no intrinsic rotation of $\hat{u}_{i}$
around the polymer axis. Thus, we define the orientation of the bound protrusions by rotating $\hat{u}_{i}$ around $\hat{t}_{i}$. Therefore, the center of a protrusion bound to chain link $k$ and rotated around the chain axis by an angle of $\gamma_{k}$
is given by:
\begin{equation}
\mathbf{r}_{\mathrm{object}}=\mathbf{r}_{k}+\left(\frac{w}{2}+R_{o}\right)\mathbf{R}(\gamma_{k},\hat{t}_{k})\hat{u}_{k}, \label{eq:r_protein}
\end{equation}
where $\mathbf{R}(\gamma_{k},\hat{t}_{k})$ is a rotation matrix by an angle $\gamma_{k}$
around $\hat{t}_{k}$ (see  Fig.~\ref{fig:illustration}). Addition of a protrusion at link $k$
slightly alters Eq.~(\ref{eq:E-hw}), requiring to test whether joint $i$
overlaps with one or more joints $0\ldots\left(i-\Delta i\right)$
and with the protrusion at $\mathbf{r}_{\mathrm{object}}$, if $i>k$.

\emph{Simulation.} We simulate the DNA chain with a bound protein using a weighted-biased sampling, Monte-Carlo approach that we used previously to simulate the configurational space of bare DNA \cite{Pollak2014}. To adapt our algorithm to the case of protein-bound DNA, we take into account not only the growing chain but also the location of the protrusion (see Eq.~(\ref{eq:r_protein})). During chain generation, upon reaching link $k$, the simulation adds a hard-wall spherical protrusion with radius $R_{o}$ at the location $\mathbf{r}_{\mathrm{object}}$. If the protrusion overlaps any of the previously-generated chain links or protrusions, the chain is discarded. After generating the configurational ensemble, we identify the subset of ``looped” chains. A chain is looped if $\mathbf{r}_{N}$ is confined to a volume $\delta\mathbf{r}$ around $\mathbf{r}_{0}$ (see Fig.~\ref{fig:illustration}), defined by:
\begin{enumerate}
        \item $d_{\mathrm{min}}\leq\left|\mathbf{r}_{N}-\mathbf{r}_{0}\right|\leq d_{\mathrm{min}}+\varepsilon$. \label{enu:loop-crit-1}
        \item $\left(\mathbf{r}_{N}-\mathbf{r}_{0}\right)$ is collinear with $\hat{u}_{0}$ within $\delta\omega'$. \label{enu:loop-crit-2}
\end{enumerate}
In our simulations, $d_{\mathrm{min}}=w$, $\varepsilon=2w$ and $\delta\omega'=2\pi\times0.1$, unless stated otherwise. Changing these parameters did not alter the results significantly, and the relatively large $\varepsilon$ chosen
minimized noise. We simulated the DNA chain with diameter $w=4.6$~nm
and Kuhn length \cite{flory_statistical_1969} $b=106$~nm, where $b$ is given by \cite{tree_is_2013}:
\begin{equation}
\frac{b}{l}=\frac{a-1+a\coth a}{a+1-a\coth a} .
\end{equation}
Chains were simulated in two stages as  compromise between resolution and running time. The first $N_1$ links of the chain were simulated with link length $l_1=0.34$~nm, corresponding to the length of a base-pair in dsDNA. We used $\Delta i\approx\frac{4}{3}\frac{w}{l}$ for this stage \cite{Brunwasser-Meirom2016}. All bound objects were positioned at links $k<N_1$. The remaining $N_2$ links of the chain were simulated with link length $l_2=w$. We denote the overall length of the chain by $L=N_{1}l_{1}+N_{2}l_{2}$, and the distance along the chain of an object binding link $k$  from the chain origin by $K=kl_1$.

For a specific choice of $\delta\mathbf{r}$, we define the probability of a polymer chain of length $L$ to form a loop, and relate it to the experimentally measurable \cite{jacobson_intramolecular_1950} Jacobson-Stockmayer factor $J(L)$  as \cite{Pollak2014}:
\begin{equation}
P_{\mathrm{looped}}\left(L\right)\equiv\int\limits _{\delta\mathbf{r}}\mathbf{C}\left(\mathbf{r}\right)\mathrm{d}\mathbf{r}
=J\left(L\right) N_{A}\sigma_{R}\delta\mathbf{r},\end{equation}
where $\mathbf{C\left(\mathbf{r}\right)}$ is the probability density function of the end-to-end vector $\mathbf{r}\equiv\mathbf{r}_{N}-\mathbf{r}_{0}$, $N_{A}$ is Avogadro's number, and $\sigma_{R}$
is the symmetry number of a polymer ring \cite{flory_statistical_1969}. In this work we study the effect of a bound object on the probability of the polymer to form a loop, with the looping criteria defined above. We quantify this effect by the looping probability ratio:
\begin{equation}
F\left(L,\mathrm{object}\right)\equiv\frac{J_{\mathrm{object}}\left(L\right)}{J_{\mathrm{baseline}}\left(L\right)}=\frac{P_{\mathrm{looped}}^{\,\mathrm{object}}\left(L\right)}{P_{\mathrm{looped}}^{\,\mathrm{baseline}}\left(L\right)}, \label{eq:F_definition}
\end{equation}
where $J_{\mathrm{baseline}}\left(L\right)$
is the J-factor for the bare polymer chain and $J_{\mathrm{object}}\left(L\right)$
is the J-factor for the polymer chain with a protrusion bound to it.

\begin{figure}[t]
        \centering{}
        \includegraphics[width=\columnwidth]{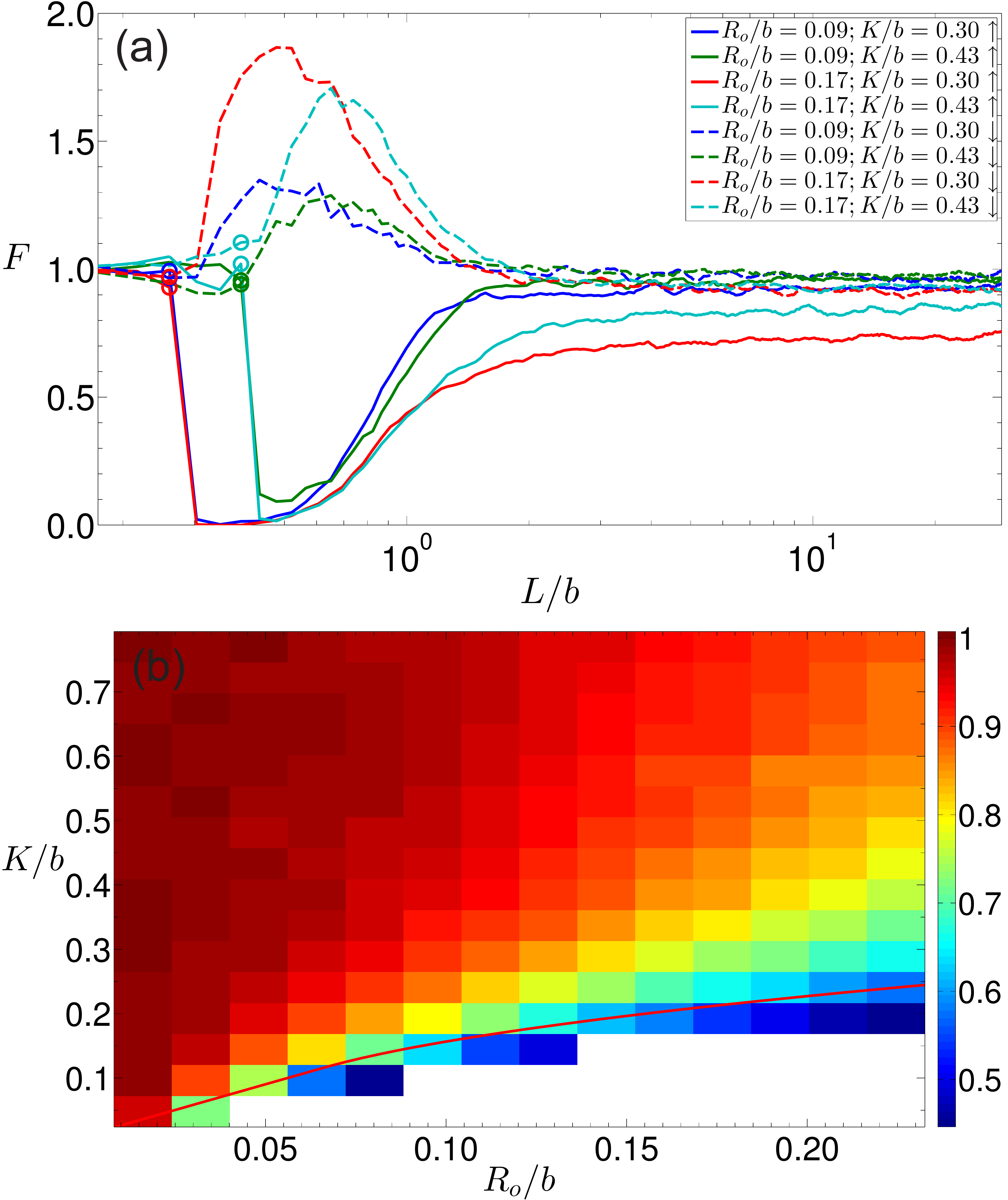}
        \caption{\label{fig:dep-size-loc-and-heatmap} Simulating looping probability ratio $F$. (a) $F$ plotted as a function of normalized chain length $\frac{L}{b}$ for several values of $R_o$ and $K$, and $\gamma_{k}=0\degree$ (solid lines) or $180\degree$ (dashed lines). The locations of the protrusions are denoted by circles on the corresponding curves. (b) $F_\infty$ plotted as a function of $R_o$ and $K$, for $\gamma_{k} = 0$. $F_{\infty}$ is calculated numerically as the average of $F\left(L\right)$ over the range of $L$ values where $F\left(L\right) \approx const$. The solid red curve is a visual aid: if the segment of the chain between the origin and the protrusion location was straight, points on the red curve would result in the protrusion touching the looping volume $\delta\mathbf{r}$.}
\end{figure}

\emph{Long-range down-regulatory effect.} We generated configurational ensembles for thick chains with a spherical protrusion of varying size $0.0434\leq R_{o}/b\leq0.217$ located a distance of $0.304\leq K/b\leq0.694$ from the chain origin along the chain, oriented either in the same direction as the looping volume $\delta\mathbf{r}$ or $180\degree$ from it (see Fig.~\ref{fig:illustration}). Plots of $F\left(L\right)$ for various values of $R_o$ and $K$ are shown in Fig.~\ref{fig:dep-size-loc-and-heatmap}(a). As we showed previously \cite{Brunwasser-Meirom2016}, in the elastic regime, protrusions positioned in-phase with $\delta\mathbf{r}$ (solid lines, \textuparrow) strongly reduce the looping probability relative to that of the bare chain, while protrusions positioned out-of-phase to $\delta\mathbf{r}$ (dashed lines, \textdownarrow) increase the looping probability. However, in the entropic regime ($L\gg b$), all chains converge to values of $F\xrightarrow{L\gg b}F_{\infty}<1$.  For protrusions with $\gamma_{k}=0$, $F_\infty$ is distinctly smaller than 1, and strongly depends on both the distance to the nearest terminus and the size of the protrusion. Conversely, for $\gamma_{k}=180\degree$, $F_\infty$ is only slightly smaller than 1, with weak dependence on both protrusion size and position. In Fig.~\ref{fig:dep-size-loc-and-heatmap}(b) we plot the value of $F_\infty$ as a function of $R_o$ and $K$, for $\gamma_{k}=0$. The figure shows a decrease in $F_\infty$ as a function of protrusion size and an increase as a function of protrusion distance from the chain origin.

\begin{figure}[t]
        \centering{}
        \includegraphics[width=\columnwidth]{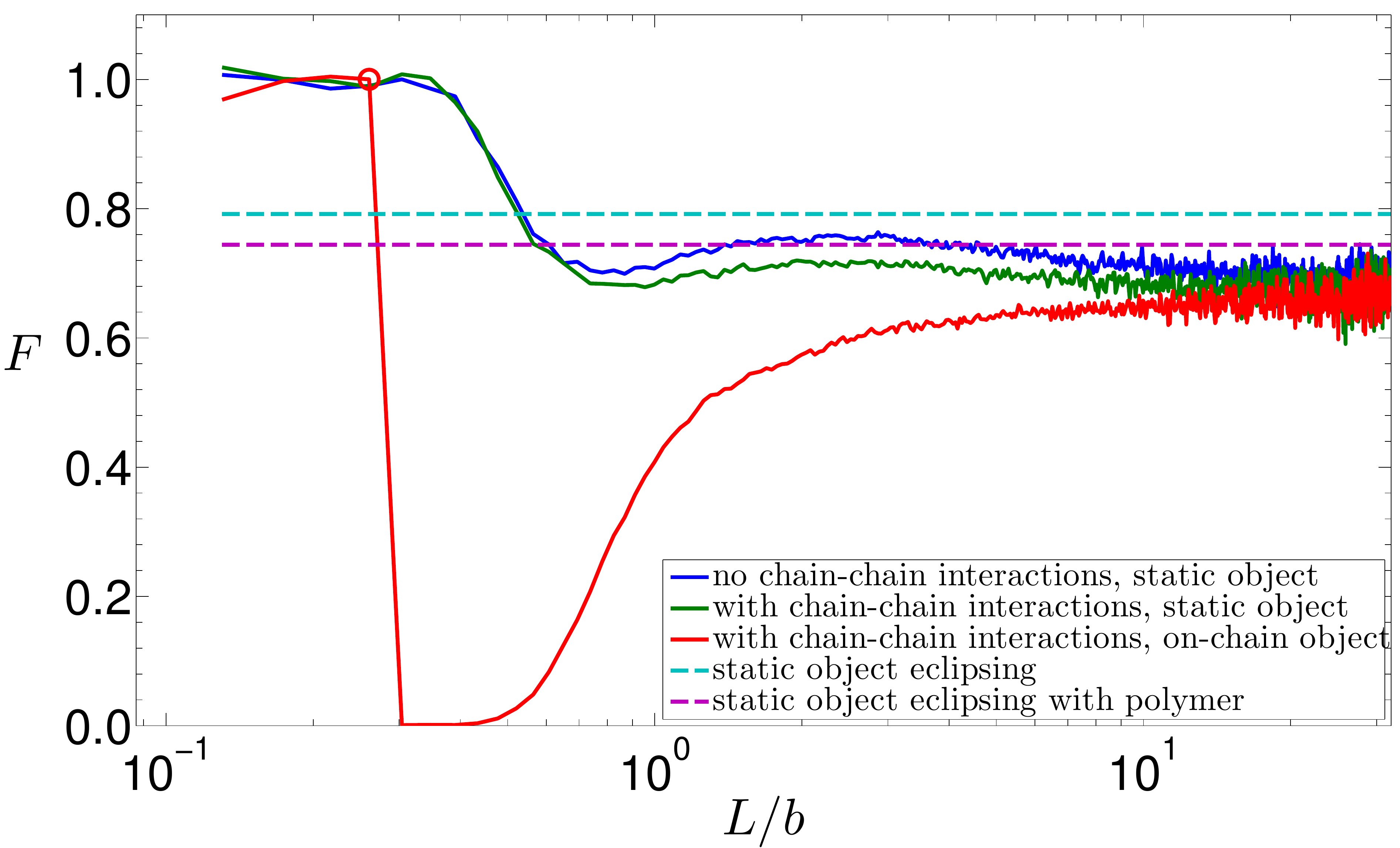}
        \caption{\label{fig:compare-eclipse-models} Simplified eclipse models. Simulation results for a chain without chain-chain interactions and a static object (solid blue line) are compared to estimates for $F_{\infty}$ of the ``rod'' (dashed cyan line) and ``terminating-segments'' (dashed magenta line) models, also for a chain without chain-chain interactions. The static object is located at $d\hat{u}_{0}$, where $d/b=$~0.391, and $R_{o}/b=$~0.217. For comparison, we plot simulation results for a chain with chain-chain interactions and the same static object parameters (solid green line), and a chain with chain-chain interaction and an on-chain object (solid read line). The on-chain object is located at $K/b$=0.303 (denoted by a circle).}
\end{figure}

\emph{``Eclipsing'' approximation.} To understand the long-range, length-independent effect shown in Fig.~\ref{fig:dep-size-loc-and-heatmap}(a), we examine the terminating segment of each looped chain of length $T\ll L$. If $T\ll b$, these segments resemble stiff rods. In the entropic regime and in the absence of protrusions, the generated ``rods'' approach $\delta\mathbf{r}$ from all directions that are unobscured by the volume of the polymer in a homogeneous fashion \cite{flory_statistical_1969}. If an object is
in close proximity to $\delta\mathbf{r}$, such that $\max\limits_{\mathbf{r}\in\delta\mathbf{r}}\left|\mathbf{r}-\mathbf{r}_{\mathrm{object}}\right|\ll T$, the object obstructs the line-of-sight of one chain terminus from the other. This eclipse-like phenomenon is manifested by a reduction in the number of polymer chains that are able to reach $\delta\mathbf{r}$. This, in turn, results in a smaller $P_{\mathrm{looped}}$ as compared with the case in which no protrusion is present. Due to the isotropy in the distribution of the chain termini orientations within $\delta\mathbf{r}$, the reduction in $P_{\mathrm{looped}}$ can be approximated by the solid angle that the eclipsing object subtends at $\delta\mathbf{r}$. Consequently, $F$ (Eq.~(\ref{eq:F_definition})) can be approximated for this ``rod model'' by:
\begin{eqnarray}
&F&_{\!\!\!\infty}\left(\tilde{\mathbf{r}}',R_{o}\right)=\left.\frac{P_{\mathrm{looped}}^{\,\mathrm{object}}\left(L,\lbrace\tilde{\mathbf{r}}',R_{o}\rbrace \right)}{P_{\mathrm{looped}}^{\,\mathrm{baseline}}\left(L\right)}\right|_{L\gg b}\notag\\ &\approx&\!\!\frac{4\pi\delta\mathbf{r}-\mathcal{I}_{\mathrm{chain}}-\mathcal{I}_{\mathrm{object}\left(\tilde{\mathbf{r}}',R_{o}\right)}+\mathcal{I}_{\mathrm{chain}\cap \mathrm{object}\left(\tilde{\mathbf{r}}',R_{o}\right)}}{4\pi\delta\mathbf{r}-\mathcal{I}_{\mathrm{chain}}}\notag\\
&=&1-\frac{\mathcal{I}_{\mathrm{object}\left(\tilde{\mathbf{r}}',R_{o}\right)}}{4\pi\delta\mathbf{r}-\mathcal{I}_{\mathrm{chain}}}+\frac{\mathcal{I}_{\mathrm{chain}\cap \mathrm{object}\left(\tilde{\mathbf{r}}',R_{o}\right)}}{4\pi\delta\mathbf{r}-\mathcal{I}_{\mathrm{chain}}}, \label{eq:F_formula-1}
\end{eqnarray}
where $R_{o}$ is the radius of the spherical object, $\tilde{\mathbf{r}}'$ is the location of the object, which could be located statically at point $\mathbf{r}'$ (in which case $\tilde{\mathbf{r}}'\equiv\mathbf{r}'$), or located on the chain a distance $K$ from the chain origin (in which case we use the terminology $\tilde{\mathbf{r}}'\equiv\tilde{\mathbf{r}}'\left(K\right)$ figuratively to specify the progression of the protrusion along the chain). $\mathcal{I}_{\mathrm{chain}}\equiv\int _{\delta\mathbf{r}}\Omega_{\mathrm{chain}}\left(\mathbf{r}\right)\mathrm{d}^{3}\mathbf{r}$, where $\Omega_{\mathrm{chain}}\left(\mathbf{r}\right)$ is the solid angle subtended at $\mathbf{r}$ by the polymer chain links. $\mathcal{I}_{\mathrm{object}\left(\tilde{\mathbf{r}}',R_{o}\right)}\equiv\int_{\delta\mathbf{r}}\Omega_{\mathrm{object}\left(\tilde{\mathbf{r}}',R_{o}\right)}\left(\mathbf{r}\right)\mathrm{d}^{3}\mathbf{r}$, where $\Omega_{\mathrm{object}\left(\tilde{\mathbf{r}}',R_{o}\right)}\left(\mathbf{r}\right)$ is the solid angle subtended at $\mathbf{r}$ by the object, and  $\mathcal{I}_{\mathrm{chain}\cap \mathrm{object}\left(\tilde{\mathbf{r}}',R_{o}\right)}$ corresponds to the solid angle contained in both $\mathcal{I}_{\mathrm{chain}}$ and $\mathcal{I}_{\mathrm{object}}$.

In order to test the eclipsing hypothesis, we first computed $F_\infty$ for the case of an object statically positioned at an off-chain location $\mathbf{r}'=d\hat{u}_{0}$, and without chain-chain interactions. In this simplified case, $F_\infty$ in Eq.~(\ref{eq:F_formula-1}) can be approximated by the following eclipsing expression:
\begin{eqnarray}
\label{eq:F_estimate_static_obj_no_chainV}
&F&_{\!\!\!\infty}\left(\mathbf{r}',R_o\right)\approx 1-\frac{\mathcal{I}_{\mathrm{object}\left(\mathbf{r}',R_o\right)}}{4\pi\delta\mathbf{r}}\\
&=&1-\frac{\int\limits_{\delta\mathbf{r}}2\pi\left[ 1-\sqrt{1-\left(\frac{R_o+w/2}{\left|\mathbf{r}-\mathbf{r}'\right|}\right)^{2}}\right]\mathrm{d}^{3}\mathbf{r}}{4\pi\delta\mathbf{r}} \equiv F^{\mathrm{static}}_{\infty},\notag
\end{eqnarray}
where we substituted
\begin{equation}
\label{eq:ugly_Omega}
\frac{\Omega_{\mathrm{object}}}{2\pi}=\negsp\negsp\!\!\!\!\int\limits _{\sqrt{1-\frac{\left(R_{o}+w/2\right)^{2}}{|\mathbf{r}-\mathbf{r}'|^{2}}}}^{1}\negsp\negsp\!\!\!\mathrm{d}\cos\theta=\!1-\sqrt{1-\frac{\left(R_{o}+w/2\right)^2}{|\mathbf{r}-\mathbf{r}'|^2}}.
\end{equation}
In Fig.~\ref{fig:compare-eclipse-models}, we compare the value computed from Eq.~(\ref{eq:F_estimate_static_obj_no_chainV}) (dashed cyan line) to $F\left(L\right)$ computed by our weighted-biased-sampling algorithm for the same conditions (solid blue line). The data show that the eclipsing approximation $F^{\mathrm{static}}_{\infty}$ overestimates $F_{\infty}$. We reasoned that the main cause for this estimation error is that Eq.~(\ref{eq:F_estimate_static_obj_no_chainV}) disregards the flexible polymer nature of the chain. We ran an additional Monte-Carlo simulation to quantify the correction resulting from polymer flexibility. Here, we generated pairs consisting of an end-terminus point in $\delta\mathbf{r}$ and a direction vector of the terminal link, both distributed uniformly. Short polymer chains of length $T$ originating at the chosen points were grown with their first links oriented in the chosen directions. These chains can be thought of as the terminating segments of long chains that have a uniform distribution of their end-termini in $\delta\mathbf{r}$. We found that the probability of a chain to overlap the object increased relative to the probability within the ``rod model'', resulting in a decrease in the probability of the chain to form a loop (magenta dashed line in Fig.~\ref{fig:compare-eclipse-models}). Using this ``terminating-segments'' correction, the discrepancy between $F_{\infty}$ from the simulation and $F^{\mathrm{static}}_{\infty}$ from Eq.~\ref{eq:F_estimate_static_obj_no_chainV} is partially accounted for. We attribute the additional reduction in the simulated $F_\infty$ to interactions between the object and the remaining $L-T$ length of the chain.

\begin{figure}[t]
        \centering{}
        \includegraphics[clip,width=\columnwidth]{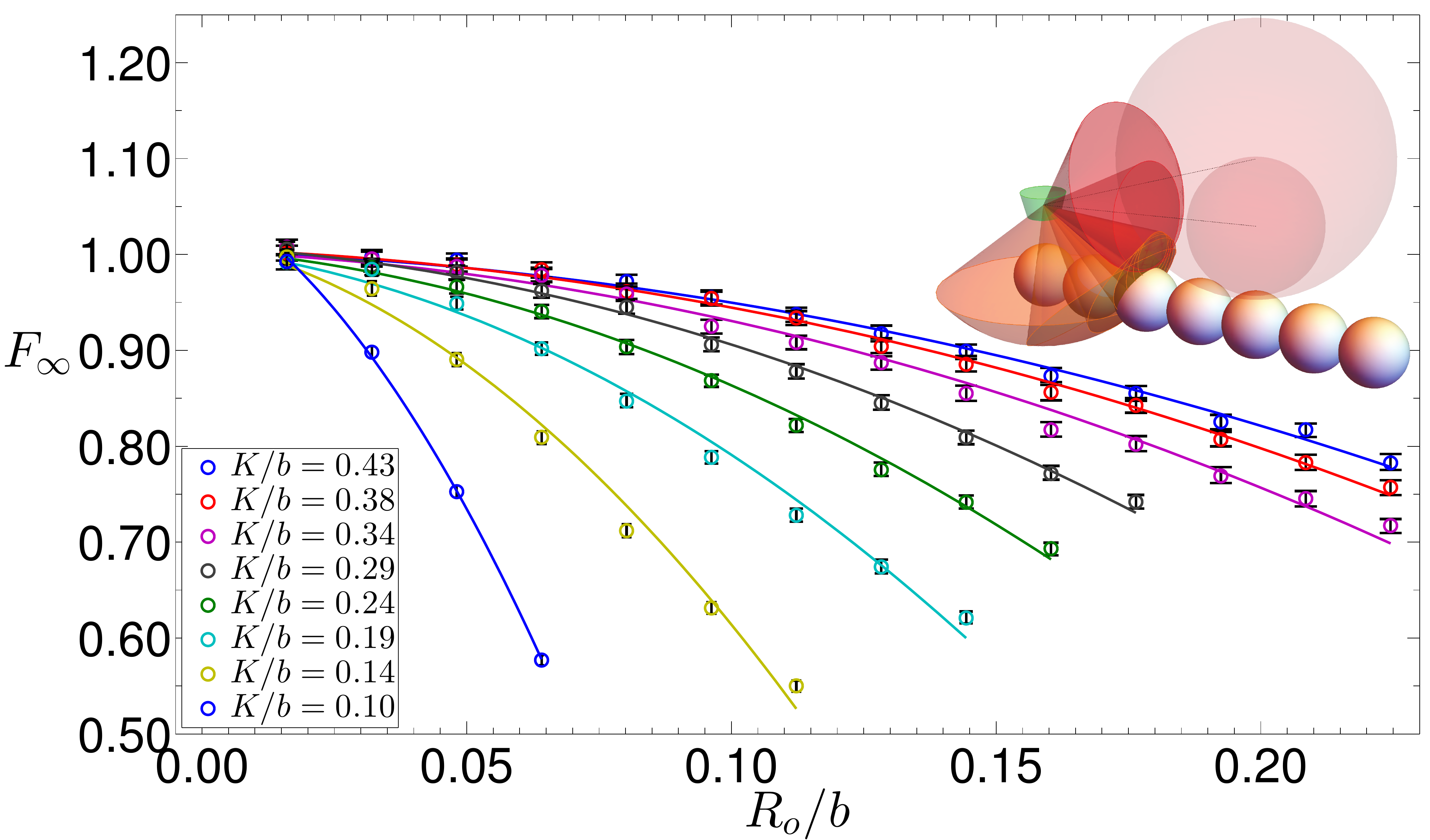}
        \caption{\label{fig:dep_on_size_location} Dependence of $F_\infty$ on protrusion size $R_{o}$. Simulation data (circles) are fit (solid curves) by $f_{K}\left(R_{o}\right)$ (Eq.~\ref{eq:fkRo}). Data points are the mean of the last $480$ points of the simulated $F$. Errorbars are $\pm$1.96 times the standard error of these points. Inset illustrates a doubling of $R_{o}$. $\delta\mathbf{r}$ is shown by the green volume. Chain links are shown by white spheres. Protrusions are shown by transparent red spheres. Solid angles subtended by the chain links and protrusion are shown by orange and red cones, respectively. $\Omega_{\mathrm{chain}}$ is the area on the unit sphere around the center of $\delta\mathbf{r}$  intersecting the orange cones. $\Omega_{\mathrm{object}\left(\tilde{\mathbf{r}}',R_{o}\right)}$ is the area on the unit sphere intersecting the red cone. $\Omega_{\mathrm{chain}\cap \mathrm{object}\left(\tilde{\mathbf{r}}',R_{o}\right)}$ is the area on the unit sphere intersecting both orange and read cones.}
\end{figure}

In Fig.~\ref{fig:dep_on_size_location} we plot $F_\infty$ as a function of an on-chain object of radius $R_o$, for several values of $K$. To compare the results of the numerical simulation to the full eclipsing model (Eq.~\ref{eq:F_formula-1}), we first note that when $K$ is kept constant, $\mathcal{I}_{\mathrm{chain}\cap \mathrm{object}\left(\mathbf{\tilde{r}}',R_{o}\right)}|_{K=const}\approx const$, as can be seen from the inset in Fig.~\ref{fig:dep_on_size_location}: the overlap between $\Omega_{\mathrm{chain}}$ (orange cones) and $\Omega_{\mathrm{object}}$ (red cones) changes only slightly when the object grows by a factor of two. Furthermore, in cases when $\left(R_{o}+w/2\right)\ll\left|\tilde{\mathbf{r}}'-\mathbf{r}\right|_{\mathbf{r}\in\delta\mathbf{r}}$, for some $\mathbf{r}\in\delta\mathbf{r}$, $\left|\tilde{\mathbf{r}}'-\mathbf{r}\right|_{K=const}$ is approximately independent of $R_{o}$. Thus, the dependence of $F_\infty$ on the radius $R_{o}$ of an on-chain object can be derived from Eq.~(\ref{eq:F_formula-1}):
\begin{eqnarray}
&F_\infty&\left(\tilde{\mathbf{r}}',R_{o}\right)|_{K=const}\approx\notag\\
&\approx&1-\frac{\mathcal{I}_{\mathrm{object}\left(\tilde{\mathbf{r}}',R_{o}\right)}}{4\pi\delta\mathbf{r}-\mathcal{I}_{\mathrm{chain}}}+\frac{\mathcal{I}_{\mathrm{chain}\cap \mathrm{object}\left(\tilde{\mathbf{r}}',R_{o}\right)}}{4\pi\delta\mathbf{r}-\mathcal{I}_{\mathrm{chain}}}\notag\\
&\approx&1-A\mathcal{I}_{\mathrm{object}\left(\tilde{\mathbf{r}}',R_{o}\right)}+B_{K}\notag\\
&\approx&1-A_{K}\left(R_{o}+w/2\right)^{2}+B_{K}\,\equiv\,f_{K}\left(R_{o}\right), \label{eq:fkRo}
 \end{eqnarray}
where we approximated $\mathcal{I}_{\mathrm{object}\left(\tilde{\mathbf{r}}',R_{o}\right)}\approx \left(R_{o}+w/2\right)^{2}\pi\int \frac{\mathrm{d}^3 \mathbf{r}}{\left|\tilde{\mathbf{r}}'-\mathbf{r}\right|^{2}}$
using Eq.~(\ref{eq:ugly_Omega})
and $\left(R_{o}+w/2\right)\ll\left|\tilde{\mathbf{r}}'-\mathbf{r}\right|_{\mathbf{r}\in\delta\mathbf{r}}$. In Fig.~\ref{fig:dep_on_size_location}(a), we fit the numerical results for different values of $K$ with functions of the form $f_{K}\left(R_{o}\right)$ (Eq.~(\ref{eq:fkRo})). The fits are in excellent agreement ($R^{2}\sim$~0.99) with the numerical data.

\begin{figure}[!t]
        \centering{}
        \includegraphics[width=\columnwidth]{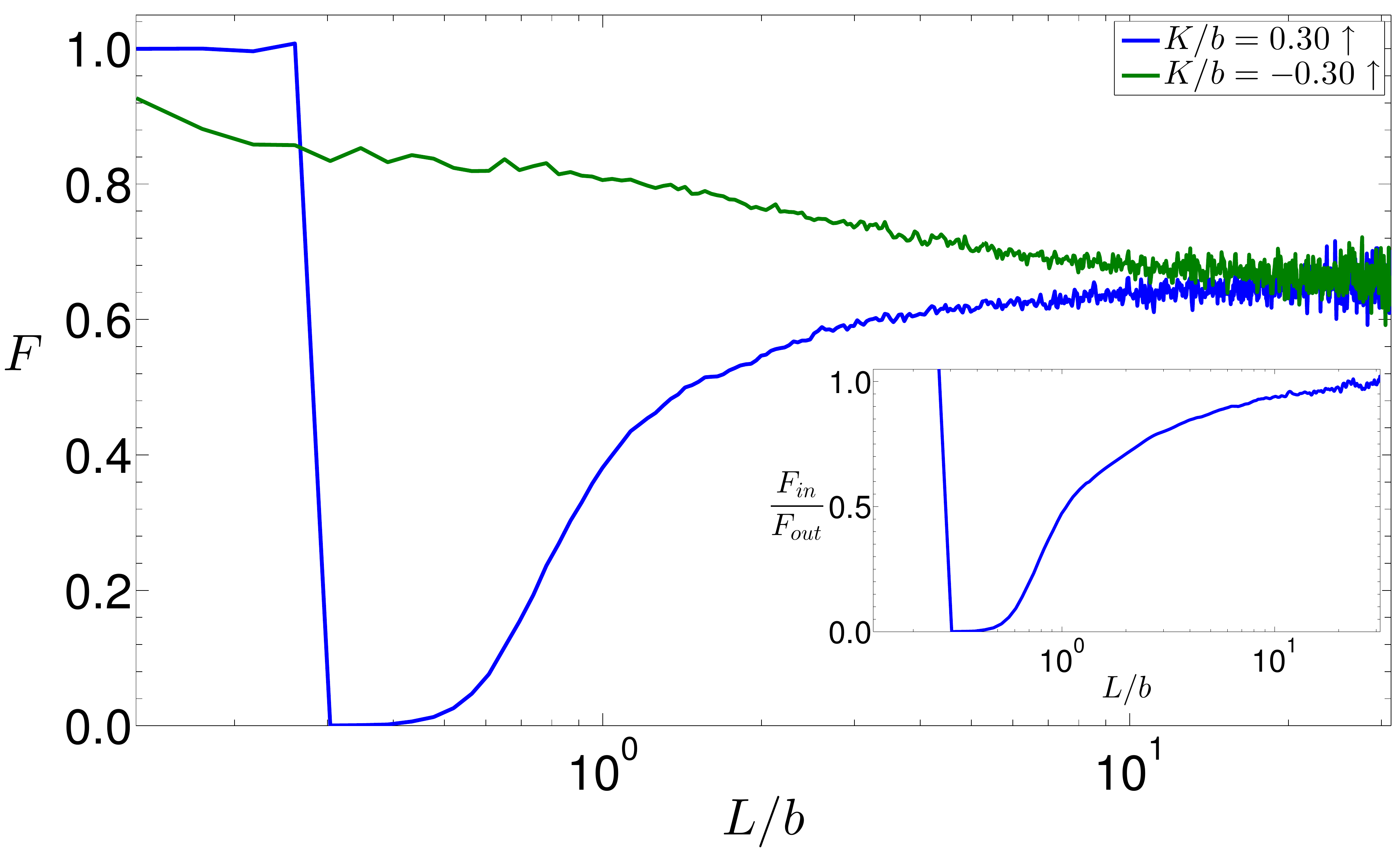}
        \caption{\label{fig:up_and_down} Dependence of $F\left(L\right)$ on location of the protrusion, either inside ($K$) or outside ($-K$) the chain segment between link 0 and the terminating link. $R_o/b=0.217$ and $K/b=\pm0.3$. Inset: the ratio $F_\infty\left(L,K\right)/F_\infty\left(L,-K\right)$.}
\end{figure}

Finally, we explored a geometry in which the protrusion was positioned at negative $K$ values. To do so, we generated an additional chain segment of length $Q$ in the direction oppsite to $\hat{t}_{1}$, starting from link 0, where $Q\gg K$. The eclipse model predicts that $F_\infty\left(\tilde{\mathbf{r}}'(K),R_{o}\right)=F_\infty\left(\tilde{\mathbf{r}}'(-K),R_{o}\right)$. We plot the results in Fig.~\ref{fig:up_and_down}. The data show that $F\left(L\right)$ for $\pm K$ initially diverge. However, for sufficiently large $L/b$, $F\left(L\right)$ for both $\pm K$ converge on the same value.

We previously established~\cite{Brunwasser-Meirom2016} that excluded-volume effects can alter the probability of looping when the chain length is on the order of the Kuhn length. The simulations and theory presented here extend this result to much longer chain lengths. In particular, our model predicts a decrease in the probability of looping that is independent of chain length for long chains in the entropic regime, provided that a sufficiently large protrusion oriented in-phase with $\delta\mathbf{r}$ is positioned within one Kuhn length of one of the chain termini. Since $F\left(L\right)$ can also be used as a measure for the biological regulatory effect induced by a protein bound to DNA in the context of looping~\cite{Brunwasser-Meirom2016}, the model presented in this Letter can be used to explain a host of natural regulatory phenomenon (e.g., ``quenching'' repression), which to date remain poorly understood.

This project received funding from the European Union’s Horizon 2020 Research And Innovation Programme under grant agreement 664918—MRG-GRammar, by the Israel Science Foundation through grant 1677/12, by the I-CORE Program of the Planning and Budgeting Committee and the Israel Science Foundation (grant 152/11). Y.P. acknowledges support provided by the Russell Berrie Nanotechnology Institute, Technion.

\bibliographystyle{apsrev4-1}
\bibliography{prl_bibliography}

\end{document}